\documentstyle[12pt]{article}
\normalsize
\def\dlta{\bigtriangleup}
\def\grd{\bigtriangledown}

\def\be{\begin{equation}}
\def\ee{\end{equation}}
\def\bea{\begin{eqnarray}}
\def\eea{\end{eqnarray}}

\newcounter{sxn}

\newcounter{axn}

\def\br{\begin{thebibliography}{abc}}
\def\er{\end{thebibliography}}

\begin{document}
\bibliographystyle{unsrt}
\footskip 1.0cm
\thispagestyle{empty}
\begin{flushright}
NSF-ITP-93-154\\
Dec. 1993\\
\end{flushright}
\vspace*{10mm}
\centerline {\Large PROPAGATION OF ELECTROMAGNETIC WAVES IN THE}
\vspace*{3mm}
\centerline {\Large  CORE OF A STRING DEFECT IN LIQUID CRYSTALS}
\vspace*{15mm}
\centerline {\large Ajit M. Srivastava and Michael Stone\footnote{
Physics Department, University of Illinois at Urbana Champaign, 1110
W. Green St. Urbana, Illinois 61801, USA}}
\vspace*{5mm}
\centerline {\it Institute for Theoretical Physics, University of California}
\centerline {\it Santa Barbara, California 93106, USA}
\vspace*{15mm}

\baselineskip=18pt

\centerline {\bf ABSTRACT}
\vspace*{8mm}

 We investigate the propagation of electromagnetic waves in the core
of a winding number one  string defect with isotropic core in
nematic liquid crystals.
We numerically solve wave equations for the TE mode and show the
existence of guided modes arising due to the variation of the
refractive index as a function of the scalar order parameter.

\newpage

\centerline {\bf 1. INTRODUCTION}
\vspace*{4mm}

 Optical fiber waveguides have been a subject of intense theoretical
and experimental investigations due to their technological importance.
Propagation of electromagnetic waves in such waveguides has been
analysed in detail for different waveguide geometries as well as for different
optical properties of the materials used. The
simplest optical fiber essentially consists of an inner
cylindrical core of a transparent medium with refractive index $n_1$
surrounded by an outer cladding of another transparent material with
refractive index $n_2$ such that $n_1 > n_2$ \cite{fiber}. There are also
graded index fibers where the refractive index decreases continuously
in radial direction \cite{grd}. Recently it has been suggested that
the propagation of electromagnetic waves in the cores of string defects
in certain transparent condensed matter systems may show interesting
behavior and that these strings may behave  as graded optical fiber
waveguides \cite{ams}.
However, the considerations in \cite{ams} were based on geometrical
optics notions which are not very appropriate for the realistic cases
when the thickness of the string defects is of the order of the wavelength
of light used.

 In this paper we analyze in detail the propagation of
electromagnetic waves in the core of a winding number one string
defects with isotropic cores in
nematic liquid crystals. Optical fibers with liquid crystalline
cores have been discussed in the literature by Lin et al. \cite{lin,lin2}
and we follow the approach used in \cite{lin} in our investigation. Our
work differs from that in \cite{lin} in two important aspects. The
propagation of light was considered in \cite{lin} for the case
when the liquid crystalline core was entirely in the nematic phase
(apart from the consideration of the effects of the propogating fields
on the order parameter as discussed in \cite{lin2}),
even though the configuration of the director field considered in \cite{lin}
corresponded to strength one line defect. The reason being that for
physically interesting wavelengths, the cylindrical cavity had to be large
making it energetically unfavorable for the strength one defect to develop
an isotropic core. We, however, consider the case when the core of the
string is in the isotropic phase. This leads to radially decreasing value
of the refractive index for the TE mode (as we will see later) resulting
in the existence of guided modes even without a transparent cladding.
This is the second difference between
our work and that in \cite{lin} where guided modes were due to
the lower value of the refractive index in the cladding material.
Even though our consideration of the isotropic core for the string restricts
us to a wavelength regime which is probably not much of practical use
(in the sense that absorption of light may be strong for such short
wavelengths), it
illustrates a novel possibility where the guided modes arise purely due to
the variation of the order parameter field. As the order parameter field
varies continuously, we expect the refractive index to do the same
(as we will explain below) making these strings into graded optical fibers.

\vspace*{8mm}
\centerline {\bf 2. WAVE PROPAGATION}
\vspace*{4mm}

  We consider a cylindrical waveguide consisting of nematic
liquid crystalline core at a temperature below the nematic-isotropic transition
temperature. The walls of the cylinder (which need not be
transparent) are assumed to be coated with appropriate material so that
the director is anchored normal to the wall leading to a strength one
string configuration \cite{tube}. We further assume that the radius of the
cylinder is small enough so that at the center of the cylinder the
order parameter vanishes leading to an isotropic core for the string. The
conditions under which this is energetically favorable have been analysed
in \cite{lin} and we refer the reader to \cite{lin} for further details.
The director then lies in the plane perpendicular to the axis of the
cylinder and points radially from the center.
We will consider the wave propagation only near the core of the string
and hence neglect any possible surface effects.

  The radial variation of the order parameter $S$ for a strength one
string defect (in the single constant approximation) is
governed by the Landau-de Gennes free energy density

\be
 F = {A \over 2} S^2 - {B \over 3} S^3 + {C \over 4} S^4 +
 {LS^2 \over 2r^2} + {L \over 2} ({\partial S \over \partial r})^2
\ee

 Here we have omitted the surface contribution as we are only interested
in the idealized case of the wave propagation near the core of the
string. Here $A = a(T - T_0)$, $T$ is the temperature and $a,~T_0,~
B,~C$ and $L$ are parameters of the material.
We take the values of these parameters
\cite{lin,par} as, $a = 0.1319 \times 10^5 J/m^3K,~B = 1.836 \times 10^5
J/m^3,~ C = 4.050 \times 10^5 J/m^3, ~ L = 2.7 \times 10^{-11} J/m$
and $T_0 = 307 K$. With these parameters, the string profile can be
obtained from the following equation corresponding to the above free
energy

\be
 {d^2S \over dr^2} + {1 \over r} {dS \over dr} -
 {S \over r^2} - {1 \over L} (A S - B S^2 + C S^3) = 0
\ee

 We have solved the above equation numerically using a
Runge-Kutta algorithm of fourth order accuracy.  The boundary conditions
for the order parameter are that $S(r)$ vanishes at $r = 0$ (corresponding to
the isotropic phase) and at large $r$, $S(r)$ approaches its asymptotic
value $\eta$ corresponding to the nematic phase. The profiles of $S(r)$
we obtain for various temperatures are in agreement with those in \cite{lin}.
We now try to determine the  spatial variation of the refractive index for
this configuration and whether it can lead to the existence of guided modes.
At the center
of the string the refractive index should have the value $n_i$ appropriate
for the isotropic phase. At large $r$, in the nematic phase, there are
two indices of refraction; $n_\Vert$ for the extra-ordinary ray and $n_\perp$
for the ordinary ray. We will consider the propagation of TE mode for
which the $n_\perp $ is the appropriate refractive index for the nematic
phase.  As we have mentioned earlier, guided TE modes
arise here due to the fact that the refractive index
decreases from the value $n_i$ at the center of the string to a smaller value
$n_\perp$ away from the center.

  We now assume that the dielectric permittivity $\epsilon_\perp$ for the
ordinary ray is proportional to the order parameter $S$ \cite{lin2}.
With the boundary conditions that $\epsilon_\perp(S = 0) = n_i^2 \epsilon_0$
and $\epsilon_\perp(S = \eta) = n_\perp^2 \epsilon_0$,
we can derive the spatial dependence
of $\epsilon_\perp(r)$ resulting from the spatial dependence of $S(r)$,

\be
 \epsilon_\perp(r) = \epsilon_0 [n_i^2 - {(n_i^2 - n_\perp^2) \over \eta}
S(r)]
\ee

where $\epsilon_0$ is the permittivity of the vacuum.

 To study the propagation of electromagnetic waves in the waveguide we
consider the Maxwell's equations
\bea
\grd \times {\bf E} & = & -{\partial{\bf B} \over \partial t} \\
\grd \times {\bf H} & = & {\partial{\bf D} \over \partial t} \\
\grd .{\bf D} & = & 0 \\
\grd .{\bf H} & = & 0
\eea
Where,
\bea
{\bf D} & = & \epsilon {\bf E} \\
{\bf B} & = & \mu_0 {\bf H}
\eea

\noindent here $\mu_0$ is the permeability of the vacuum (we will assume
this to be the same for the medium as well) and $\epsilon$ is the
permittivity of the medium.

  We consider the propagating modes in the following form

\be
{\bf E}(r,z,\phi) = {\bf E}(r) e^{i(\omega t - \beta z)}
\ee

\noindent with similar $(z,t)$ dependence for ${\bf H}$ as well. $\omega$ and
$\beta$ are the optical frequency and propagation constants
respectively. Uncoupled equations for field components can be obtained From
Eqns.(4)-(9) for the TE mode by taking ${\bf E} = (0, E_\phi, 0)$ and
${\bf H} = (H_r, 0, H_z)$. We assume that the director varies uniformly
with the azimuthal angle. Since $E_\phi$ is always normal to the director,
the permittivity $\epsilon_\perp$ only depends on $r$ and is given by Eqn.(3).
We get,

\be
{\partial^2 E_\phi \over \partial r^2} + {1 \over r}{\partial E_\phi
\over \partial r} + [\omega^2 \mu_0 \epsilon_0 \{ n_i^2 - {(n_i^2 -
n_\perp^2) \over \eta} S(r) \}
- \beta^2 - {1 \over r^2}] E_\phi =  0
\ee
\vspace*{-10mm}
\bea
H_r & = & -{\beta \over \omega \mu_0} E_\phi \\
H_z & = & {i \over \omega \mu_0} ({\partial E_\phi \over \partial r}
+ {E_\phi \over r})
\eea

 Here we have replaced $\epsilon$ by $\epsilon_\perp$ from Eqn.(3)
as appropriate for the TE mode. Using $\epsilon_0 \mu_0 \omega^2
= (2 \pi / \lambda)^2$ and writing $\beta$ as $2\pi/\lambda_z$,
we can rewrite the above equation as

\be
{\partial^2 E_\phi \over \partial r^2} + {1 \over r}{\partial E_\phi
\over \partial r} + [A - B S(r) - {1 \over r^2}] E_\phi =  0
\ee
\vspace*{-10mm}
\bea
A & = & 4\pi^2 [{n_i^2 \over \lambda^2} - {1 \over \lambda_z^2}] \\
B & = & {4 \pi^2 \over \eta \lambda^2} [n_i^2 - n_\perp^2]
\eea

 We may mention here that guided TM modes will not be expected to arise for
the case under consideration.
This is because for such modes  $E_r$ does not vanish and for $E_r$ the
corresponding refractive index $n_\Vert$ in the nematic phase, away from
the string center, is larger than the value $n_i$ at the center.

\vspace*{8mm}
\centerline {\bf 3. NUMERICAL SOLUTIONS}
\vspace*{4mm}

We solve Eqn.(2) for various values
of temperatures and use the resulting profile of $S(r)$  in the expression
of $\epsilon_\perp(r)$ to solve for $E_\phi$ in Eqn.(14). Numerical solutions
are again obtained by using a Runge-Kutta algorithm of fourth order accuracy.
Boundary
conditions for $E_\phi$ are that $E_\phi = 0$ at $r = 0$ and at $r
\rightarrow \infty$ (for guided mode). We choose a fixed value of $B$
and vary $A$ to find the solution of Eqn. (14) which is well
behaved at large $r$. We then decrease the value of $B$ to search for
the largest value of the wavelength $\lambda$ for which a guided mode
can be found. Values of $\lambda$ and $\lambda_z$ can be obtained from
the values of $A$ and $B$ from Eqns. (15)-(16). Profiles of $H_r$ and $H_z$
are obtained from Eqns(12)-(13).

We take the values of the indices of refraction as,
$n_i = 1.6$ and $n_\perp$ = 1.5 \cite{lin}. It is important to mention
here that these values of refractive indices are typical at optical
wavelengths while the wavelengths relevent for our case come out to be
much shorter. Though, we note that numerically we obtain
the value of $A$ and $B$ and the refractive indices are used only to
determine the  corresponding values of $\lambda$ and
$\lambda_z$ from Eqns.(15)-(16). We take the above values of refractive
indices as an example. Thus, if the value of $n_i^2 - n_\perp^2$
is larger for shorter wavelengths then the corresponding wavelength will
also be larger. [As we mentioned earlier, for short wavelengths the
absorption of light may also be strong. We will, however, not worry about
this. Our main aim here is to illustrate the qualitative feature of
the existence of guided modes purely due to the spatial variation of the
order parameter.]

Fig. (1) shows the plots of $E_\phi$ and $H_z$ for different temperatures
and for different values of $\lambda$.  $H_r$ can be simply obtained from
Eqn.(13). Our choice of temperatures was guided by the
consideration of finding largest values of the core radius
of the string which would lead to large, possibly practically more
interesting,  values for the wavelength $\lambda$.
The most favorable value of $\lambda$ we could find
for our set of parameters is about 2670 $\AA$ corresponding to the plot
shown in Fig. 1a.

\vspace*{8mm}
\centerline {\bf 5. CONCLUSIONS}
\vspace*{4mm}

 We have shown the existence of guided modes for a strength one string
configuration with isotropic core due to the variation of the
refractive index as a function of the order parameter. Even though
the consideration of isotropic core has restricted as to the consideration
of wavelengths of the order of 2600 $\AA$, we believe that our results
point out an interesting possibility that for suitable liquid crystal samples
for which the string size can be made larger, guided modes can be found
purely within the liquid crystalline core as compared to the cases where
the lower refractive index of the cladding is responsible for guided modes.
Further, due to continuous dependence of the refractive index on the
order parameter, the resulting waveguide is similar to a graded  index
fiber.

\vskip .3in
\centerline {\bf ACKNOWLEDGEMENTS}
\vskip .1in

 We would like to thank M. Lee and P. Palffy-Muhoray for providing
ref. \cite{par}. This work was supported by the
National Science Foundation under grant numbers PHY89-04035 and DMR91-22385.

\newpage

\vskip .3in
\centerline {\bf FIGURE CAPTIONS}
\vskip .1in

(1) Profile of the order parameter $S(r)$ is shown by the solid curve while
$E_\phi$ and ${1 \over 2}H_z$ are shown by the dashed and the dotted curves
respectively. (a) Plots for $\dlta T = 1.44$ and $\lambda = 2677 \AA$.
$\lambda_z$ is found to be = 1790 $\AA$. (b) Plots for $\dlta T = 1.35$,
$\lambda = 1420 \AA$ and $\lambda_z = 945 \AA$.

\end{document}